\documentstyle[11pt,psfig,twoside]{article}

\begin{document}

\newcommand{\dd}{\,{\rm d}}
\newcommand{\ie}{{\it i.e.},\,}
\newcommand{\etal}{{\it et al.\ }}
\newcommand{\eg}{{\it e.g.},\,}
\newcommand{\cf}{{\it cf.\ }}
\newcommand{\vs}{{\it vs.\ }}
\newcommand{\zdot}{\makebox[0pt][l]{.}}
\newcommand{\up}[1]{\ifmmode^{\rm #1}\else$^{\rm #1}$\fi}
\newcommand{\dn}[1]{\ifmmode_{\rm #1}\else$_{\rm #1}$\fi}
\newcommand{\upd}{\up{d}}
\newcommand{\uph}{\up{h}}
\newcommand{\upm}{\up{m}}
\newcommand{\ups}{\up{s}}
\newcommand{\arcd}{\ifmmode^{\circ}\else$^{\circ}$\fi}
\newcommand{\arcm}{\ifmmode{'}\else$'$\fi}
\newcommand{\arcs}{\ifmmode{''}\else$''$\fi}
\newcommand{\MS}{{\rm M}\ifmmode_{\odot}\else$_{\odot}$\fi}
\newcommand{\RS}{{\rm R}\ifmmode_{\odot}\else$_{\odot}$\fi}
\newcommand{\LS}{{\rm L}\ifmmode_{\odot}\else$_{\odot}$\fi}

\newcommand{\Abstract}[2]{{\footnotesize\begin{center}ABSTRACT\end{center}
\vspace{1mm}\par#1\par
\noindent
{~}{\it #2}}}

\newcommand{\TabCap}[2]{\begin{center}\parbox[t]{#1}{\begin{center}
  \small {\spaceskip 2pt plus 1pt minus 1pt T a b l e}
  \refstepcounter{table}\thetable \\[2mm]
  \footnotesize #2 \end{center}}\end{center}}

\newcommand{\TableSep}[2]{\begin{table}[p]\vspace{#1}
\TabCap{#2}\end{table}}

\newcommand{\FigCap}[1]{\footnotesize\par\noindent Fig.\  %
  \refstepcounter{figure}\thefigure. #1\par}

\newcommand{\TableFont}{\footnotesize}
\newcommand{\TableFontIt}{\ttit}
\newcommand{\SetTableFont}[1]{\renewcommand{\TableFont}{#1}}

\newcommand{\MakeTable}[4]{\begin{table}[htb]\TabCap{#2}{#3}
  \begin{center} \TableFont \begin{tabular}{#1} #4 
  \end{tabular}\end{center}\end{table}}

\newcommand{\MakeTableSep}[4]{\begin{table}[p]\TabCap{#2}{#3}
  \begin{center} \TableFont \begin{tabular}{#1} #4 
  \end{tabular}\end{center}\end{table}}

\newenvironment{references}%
{
\footnotesize \frenchspacing
\renewcommand{\thesection}{}
\renewcommand{\in}{{\rm in }}
\renewcommand{\AA}{Astron.\ Astrophys.}
\newcommand{\AAS}{Astron.~Astrophys.~Suppl.~Ser.}
\newcommand{\ApJ}{Astrophys.\ J.}
\newcommand{\ApJS}{Astrophys.\ J.~Suppl.~Ser.}
\newcommand{\ApJL}{Astrophys.\ J.~Letters}
\newcommand{\AJ}{Astron.\ J.}
\newcommand{\IBVS}{IBVS}
\newcommand{\PASP}{P.A.S.P.}
\newcommand{\Acta}{Acta Astron.}
\newcommand{\MNRAS}{MNRAS}
\renewcommand{\and}{{\rm and }}
\section{{\rm REFERENCES}}
\sloppy \hyphenpenalty10000
\begin{list}{}{\leftmargin1cm\listparindent-1cm
\itemindent\listparindent\parsep0pt\itemsep0pt}}%
{\end{list}\vspace{2mm}}

\def\TYLDA{~}
\newlength{\DW}
\settowidth{\DW}{0}
\newcommand{\dw}{\hspace{\DW}}

\newcommand{\refitem}[5]{\item[]{#1} #2%
\def\REFARG{#3}\ifx\REFARG\TYLDA\else, {\it#3}\fi
\def\REFARG{#4}\ifx\REFARG\TYLDA\else, {\bf#4}\fi
\def\REFARG{#5}\ifx\REFARG\TYLDA\else, {#5}\fi.}

\newcommand{\Section}[1]{\section{#1}}
\newcommand{\Subsection}[1]{\subsection{#1}}
\newcommand{\Acknow}[1]{\par\vspace{5mm}{\bf Acknowledgments.} #1}
\pagestyle{myheadings}

\def\thefootnote{\fnsymbol{footnote}}

\begin{center}
{\Large\bf 
Variable Stars in the Archival HST Data
of Globular Clusters M13, M30 and NGC~6712}
\vskip1cm
{\bf
Pawe\l\ ~~P~i~e~t~r~u~k~o~w~i~c~z~~and~~ Janusz ~~K~a~l~u~z~n~y\\}
\vskip3mm
{Copernicus Astronomical Center, Bartycka 18, 00-716 Warsaw, Poland\\
e-mail: (pietruk,jka)@camk.edu.pl}
\end{center}

\Abstract{

We have analyzed archival {\it HST/WFPC2} time-series data of the
central parts of globular clusters M13, M30 and NGC~6712
in search of variable objects. Among a total of 21 identified
variables there are 15 new discoveries. The sample includes
nine RR~Lyr stars, two SX~Phe stars  and seven W~UMa-type contact
binaries. One object is preliminarily classified as a detached
eclipsing binary and another as an ellipsoidal variable.
}
{globular clusters: individual: M13, M30, NGC~6712 -- binaries: eclipsing
-- stars: variables: RR Lyr}

\Section{Introduction}

The {\it Hubble Space Telescope} has an excellent resolving power
and therefore offers an exeptional opportunity to explore the
crowded central regions of Galactic globular clusters.
The centers represent rich environments for the study of
stellar evolution, dynamical processes in stellar systems 
as well as evolution of close binary systems.

Albrow \etal (2001) used HST/WFPC2 time-series data
to identify over one hundred variables in the central part 
of 47 Tuc. Pritzl \etal (2003) analyzed the archival data on NGC 6441 
and located  57 variables,  of which 38 were RR Lyr stars.
Similarly, Pietrukowicz \& Kaluzny (2003) used the archival WFPC2 
images to detect 8 new variables  in the central region of 
globular cluster M22.

In this contribution we report indentification of several new
variable stars in fields covering nuclear regions of globular 
clusters  M13, M30 and NGC~6712.

\section{Data Reductions and Results}

Observational data consisting of processed WFPC2 images were obtained 
from the  Multimission Archive at Space Telescope 
\footnote{http://archive.stsci.edu}.
A condensed log of observations used is given in Table 1.

{\footnotesize
\begin{table}
\begin{center}
\caption{\small Data sets}

\vspace{0.4cm}

\begin{tabular}{lclcc}
\hline
Name     & Program & Date              & Filter & Exposures \\
\hline
M13      & GO 8278 & 1999 Nov 10        & F555W & $25 \times 80$ s \\
         &         & 1999 Nov 10        & F814W & $25 \times 140$ s \\
M30      & GO 7379 & 1999 May 31--Jun 1 & F336W & $8 \times 200$ s, $20 \times 300$ s, $10 \times 400$ s \\
         &         & 1999 May 31--Jun 1 & F555W & $32 \times 23$ s \\
         &         & 1999 May 31--Jun 1 & F814W & $28 \times 30$ s \\
NGC 6712 & GO 6121 & 1995 May 25--26    & F300W & $8 \times 260$ s, $44 \times 300$ s, $1 \times 400$ s \\
         &         & 1995 May 28        & F336W & 160 s \\
         &         & 1995 May 28        & F439W & 160 s \\
         &         & 1995 May 28        & F555W & 60 s \\
         &         & 1995 May 28        & F675W & 60 s \\
         &         & 1995 May 28        & F814W & 120 s \\
\hline
\end{tabular}
\end{center}
\end{table}
}

Data reductions and analysis were performed using an approach which is
described in more detail in Pietrukowicz \& Kaluzny (2003).
Profile photometry was extracted with the help of the HSTphot
package (Dophin 2000a; 2000b). The same package was used to
transform instrumental  photometry to the standard $UBVR_{C}I_{C}$
system. The search for variable stars was performed with the TATRY
code using the multi-harmonic periodogram of Schwarzenberg-Czerny (1996). 
Periodograms were calculated for periods ranging from 
0.01 day to the time span of a given data set. A total of 21 variable stars
were detected. In Table 2 we provide the positional data sufficient for their
unambiguous identification.

\subsection{M13 Variables}

{\footnotesize
\begin{table}
\begin{center}
\caption{Equatorial coordinates and ($X,Y$) positions of variables
on the HST/WFPC2 images}

\vspace{0.4cm}

\begin{tabular}{lccccc}
\hline
Name       &        RA(2000.0)         &         Dec(2000.0)          & Chip & Location & Dataset name \\
 & & & & ($X,Y$) & \\
\hline
\hline
M13\_01     & 16\uph41\upm41\zdot\ups35 & 36\arcd27\arcm04\zdot\arcs6 & WF2 & (119,163) & u5bt0104r \\
M13\_02     & 16\uph41\upm38\zdot\ups77 & 36\arcd26\arcm20\zdot\arcs9 & WF3 & (505,401) & u5bt0104r \\
M13\_03     & 16\uph41\upm42\zdot\ups89 & 36\arcd27\arcm00\zdot\arcs3 & WF2 & (287,259) & u5bt0104r \\
M13\_04     & 16\uph41\upm42\zdot\ups96 & 36\arcd27\arcm27\zdot\arcs3 & PC1 & (104,758) & u5bt0104r \\
\hline
NGC6712\_01 & 18\uph53\upm04\zdot\ups83 & ~-8\arcd42\arcm20\zdot\arcs9 & PC1 & (422,434) & u2of0101t \\
NGC6712\_02 & 18\uph53\upm04\zdot\ups12 & ~-8\arcd42\arcm04\zdot\arcs2 & PC1 & ( 54,671) & u2of0101t \\
NGC6712\_03 & 18\uph53\upm05\zdot\ups98 & ~-8\arcd41\arcm33\zdot\arcs9 & WF2 & ( 53,325) & u2of0101t \\
NGC6712\_04 & 18\uph53\upm03\zdot\ups18 & ~-8\arcd41\arcm39\zdot\arcs7 & WF2 & (471,269) & u2of0101t \\
NGC6712\_05 & 18\uph53\upm04\zdot\ups33 & ~-8\arcd42\arcm27\zdot\arcs3 & PC1 & (565,598) & u2of0101t \\
NGC6712\_06 & 18\uph53\upm06\zdot\ups80 & ~-8\arcd41\arcm31\zdot\arcs7 & WF3 & (355,165) & u2of0101t \\
NGC6712\_07 & 18\uph53\upm08\zdot\ups83 & ~-8\arcd42\arcm31\zdot\arcs3 & WF4 & (468,320) & u2of0101t \\
NGC6712\_08 & 18\uph53\upm08\zdot\ups07 & ~-8\arcd40\arcm52\zdot\arcs0 & WF3 & (756,353) & u2of0101t \\
NGC6712\_09 & 18\uph53\upm04\zdot\ups10 & ~-8\arcd42\arcm25\zdot\arcs5 & PC1 & (526,672) & u2of0101t \\
\hline
M30\_01     & 21\uph40\upm24\zdot\ups25 & ~-23\arcd11\arcm47\zdot\arcs1 & WF4 & (401,590) & u5fw0106r \\
M30\_02     & 21\uph40\upm21\zdot\ups80 & ~-23\arcd10\arcm50\zdot\arcs8 & PC1 & (500,530) & u5fw0101r \\
M30\_03     & 21\uph40\upm22\zdot\ups14 & ~-23\arcd10\arcm44\zdot\arcs3 & PC1 & (325,511) & u5fw0101r \\
M30\_04     & 21\uph40\upm22\zdot\ups02 & ~-23\arcd10\arcm46\zdot\arcs7 & PC1 & (388,515) & u5fw0101r \\
M30\_05     & 21\uph40\upm22\zdot\ups21 & ~-23\arcd10\arcm48\zdot\arcs3 & PC1 & (392,448) & u5fw0101r \\
M30\_06     & 21\uph40\upm22\zdot\ups51 & ~-23\arcd10\arcm54\zdot\arcs4 & PC1 & (465,303) & u5fw0101r \\
M30\_07     & 21\uph40\upm23\zdot\ups28 & ~-23\arcd10\arcm40\zdot\arcs6 & PC1 & ( 85,247) & u5fw0101r \\
M30\_08     & 21\uph40\upm22\zdot\ups51 & ~-23\arcd10\arcm50\zdot\arcs9 & PC1 & (397,342) & u5fw0101r \\
\hline
\end{tabular}
\end{center}
\end{table}
}

{\footnotesize
\begin{table}
\begin{center}
\caption{Photometric data for the M13 variables}

\vspace{0.4cm}

\begin{tabular}{lccccll}
\hline
Name & $V_{max}$ & $\Delta V$ & $I_{C,max}$ & $\Delta I_C$ & $P$ [d] & Type \\
\hline
M13\_01 & 17.12 & 0.18 & 16.71 & 0.11 & 0.0535(4) & SX \\
M13\_02 & 17.01 & 0.20 & 16.58 & 0.22 & 0.0644(5) & SX \\
M13\_03 & 19.96 & 0.44 & 19.12 & 0.48 & 0.223(1)  & EW \\
M13\_04 & 19.15 & 0.19 & 18.39 & 0.18 &   -       & EA \\
\hline
\end{tabular}
\end{center}
\end{table}
}

Globular cluster M13=NGC~6205 was monitored by the HST/WFPC2 for a period 
spanning approximately 7 hours. Fig. 1 displays the {\it rms} deviation as
a function of average magnitude in $I_C$ and $V$ bands for the light curves
of 23374 and 21250 stars, respectively.

The most recent list of variable stars from the  central region of M13
was published by Kopacki \etal (2003). Among a total of 26
variables in that list 13 turn out to be located in the investigated WFPC2
area. However, due to severe saturation of their images no useful 
photometry could be extracted for any of these stars.
Our search for variability led to detection of four new variables.
Fig. 2 shows their $I_C$ and $V$ light curves
while Fig. 3 displays their location on the $V/V-I$ color-magnitude diagram. 
Table 3 lists basic photometric data for new variables. 
Objects M13\_01 and M13\_02 can be securely classified as SX~Phe stars.
Both of them belong to the population of cluster blue stragglers
what is a rule for SX~Phe observed in globular clusters. 
The light curve and period of M13\_03 indicates that it is a  W~UMa-type 
contact binary. The variable is located slightly above the main-sequence of
its parent cluster what is consistent with its binary nature.\\ 
Star M13\_04 exhibited an eclipse-like event at the end of observations. 
The drop of luminosity amounted to about 0.2 mag in both employed filters.
Preliminary classification of the object as an eclipsing detached 
binary is consistent with its location above the cluster main-sequence 
in the color-magnitude diagram.

The region marked with a quadrangle in Fig. 3 includes 27
candidates blue stargglers. Light curves of these stars were 
examined individually. None of them, besides two alredy discussed 
objects, showed any evidence for a short period variability with 
an amplitude exceeding 0.02 mag. 
   
\subsection{NGC~6712 Variables}

The HST was pointed continuously at the globular cluster
NGC~6712 (C~1850-087) for a period of 13.5~h on 1995 May 25--26.
Some gaps in the data occured during Earth occultations and 
during South Atlantic Anomaly passages. Time-series observations 
consisting of 53 exposures were collected in the wide-band F300W filter. 
Additional exposures in F336W, F439W, F555W, F675W and F814W filters
were taken over a  period of 20 min on 1995 May 28.
The quality of  photometry derived for  the F300W filter is 
illustrated in Fig. 4, where we show the {\it rms} of light curves 
\vs the average magnitude for a total of 4413 of analyzed stars.

Nine periodic variables were detected. Table 4 lists some of their basic 
photometric characteristics based on observations in the F300W filter.
Phased light curves are presented in Fig. 5.
The color-magnitude diagrams of the cluster with marked  positions 
of variables are shown in Fig. 6. The adopted colors are based
on pairs of frames separated by no more than 6 minutes. Note also that 
magnitudes and colors plotted in Fig. 6 were measured
at some random phases of variables.\\  
Object \#1 is a well known optical counterpart of the low mass X-ray binary 
X1850-087 (Homer \etal 1996). Our objects \#2, \#3 and \#4 correspond to
variables V20, V12 and V19  from Clement \etal (2001), respectively. 
We provide a revised value of period for the variable V20. Using the HST
photometry we derived $P=0.249\pm 0.003$~d. The period of 0.330870~d
listed in Clement \etal (2001) does not fit the HST data. 
Moreover, reanalysis of photometry published by  Sandage, Smith \& Norton 
(1966) allows to refine the period of V20 to $P=0.248489\pm 0.000002$~d.  
Variability of RR Lyr star \#5 has been recently reported by Tuairisg 
et al. (2003). Our photometry confirms the period derived by that 
group for the variable. Stars \#6--9 are new identifications. 
Objects \#6 and \#7 belong to RR Lyr variables of RRc subtype. 
We note that star \#7 exhibits unstable light curve with some cycle-to-cycle 
changes visible in obtained photometry.
Objects  \#8 and \#9 are preliminarly classified as W~UMa-type contact 
binaries. Such classification is consistent with shape of observed
light curves and with location of these variables on the color-magnitude 
diagram of the cluster. In particular, variable \#8 is a candidate 
for cluster blue straggler.

The region marked with a quadrangle in Fig. 6 includes 39          
candidates blue stargglers. Light curves of these stars were
examined individually. None of them, besides one alredy discussed 
object, showed any evidence for a short period variability with
an amplitude exceeding 0.02 mag.

{\footnotesize
\begin{table}
\begin{center}
\caption{Photometric data for the NGC 6712 variables}

\vspace{0.4cm}

\begin{tabular}{lccll}
\hline
Name & F300W$_{max}$ & F300W$_{min}$ & $P$ [d]  & Type \\
\hline
NGC6712\_01 & 19.46 & 19.65 & 0.01429(3)  & LMXB \\
NGC6712\_02 & 17.78 & 18.26 & 0.249(3)    & RRc       \\ 
NGC6712\_03 & 17.76 & 18.60 & 0.502776   & RRab      \\ 
NGC6712\_04 & 17.50 & 18.01 & 0.423900   & RRc       \\ 

NGC6712\_05 & 17.61 & 18.39 & 0.58(6)    & RRab      \\
NGC6712\_06 & 17.58 & 17.75 & 0.247(3)   & RRc       \\
NGC6712\_07 & 18.62 & 19.05 & 0.211(3)   & RRc       \\
NGC6712\_08 & 20.12 & 20.66 & 0.302(5)   & EW        \\
NGC6712\_09 & 21.16 & 21.88 & 0.306(10)  & EW        \\
\hline
\end{tabular}
\end{center}
\end{table}
}

\subsection{M30 Variables}

{\footnotesize
\begin{table}
\begin{center}
\caption{Photometric data for the M30 variables}

\vspace{0.4cm}

\begin{tabular}{lccccccll}
\hline
Name & $U_{max}$ & $\Delta U$ & $V_{max}$ & $\Delta V$ & $I_{C,max}$ & $\Delta I_C$ & $P$ [d] & Type \\
\hline
M30\_01 & 14.96 & 0.92 &   -   &   -  &   -   &   -  & 0.751(9)  & RRab    \\ 
M30\_02 & 14.94 & 1.10 & 15.21 & 0.23 & 14.13 & 0.66 & 0.689(8)  & RRab    \\
M30\_03 & 15.26 & 0.52 & 15.08 & 0.30 & 14.54 & 0.32 & 0.341(1)  & RRc     \\
M30\_04 & 18.01 & 0.26 & 17.90 & 0.27 & 17.51 & 0.24 & 0.284(1)  & EW      \\
M30\_05 & 17.39 & 0.37 & 17.28 & 0.28 & 17.04 & 0.32 & 0.3175(12)& EW      \\
M30\_06 & 19.58 & 0.57 & 19.54 & 0.54 & 18.82 & 0.52 & 0.211(1)  & EW      \\
M30\_07 & 20.38 & 0.83 & 20.06 & 0.75 & 19.20 & 0.73 & 0.2149(5) & EW      \\
M30\_08 & 19.71 & 0.43 & 19.72 & 0.33 & 19.20 & 0.36 & 0.386(3)  & Ell     \\
\hline
\end{tabular}
\end{center}
\end{table}
}

The central part of the globular cluster M30=NGC~7099  was observed in 
three filters for a period spanning 30\zdot\uph5.
The quality and depth of derived photometry can be inferred from Fig. 7. 
Light curves for a total of 11096, 10578 and 6833 stars were analysed
for filters $I_C$, $V$ and $U$, respectively.

Eight cetrain periodic variables were detected.
Table 5 lists some of their basic photometric characteristics
while phased light curves are shown in Fig. 8. The $U/U-V$ and $V/V-I$
color-magnitude diagrams of the cluster with positions of variables
marked are presented in Fig. 9. It is remarkable that 8 out of 9 
variables were detected on the PC1 chip which covers nucleus of the cluster
but contains only about 32\% of analyzed stars.

Object M30\_01 corresponds to variable V1 from the catalogue of Clement 
\etal (2001).  That star could be measured only in the $U$ band but
obtained light curve can be phased with the period listed in 
Clement \etal (2001). The 8 remaining variables are new identifications. 
The variables \#2 and \#3 belong to RRab and RRc stars, respectively. 
Based on observed periods and shape of light curves we have classified
stars \#4--7 as W~UMa-type contact binaries. Such classification is 
further supported by observed location of these objects on the cluster 
color-magnitude diagram. Objects \#4 and \#5 are located among cluster blue
stragglers while objects \#6 and \#7 occupy positions near the red edge of 
cluster main-sequence.

Object \#8 also shows variability with a period falling formally 
within a range  of periods observed for W~UMa-type binaries. However, its 
light curve  exhibits too sharp maxima (or alternatively too wide minima) for
a contact binary. Also observed location of the variable on the 
cluster color-magnitude diagram would be atypical for contact binaries
detected so far in globular clusters.
The light curve of star \#8 exhibits two minima of slightly different depth.
One may also note that extrema of the light curve are separated 
by about $0.5~P$.  We propose that object \#8
is a close binary belonging to ellipsoidal variables although we cannot 
exclude the possibility that it is an eclipsing binary.

The region marked with a quadrangle in Fig. 9 includes 65
candidates blue stargglers. Light curves of these stars were
examined individually. None of them, besides two alredy discussed 
objects, showed any evidence for a short period variability with
an amplitude exceeding 0.02 mag.   
We also failed to detect any variables
among stars located within 1\zdot\arcs0 from the nominal positions of
the binary millisecond pulsar M30A recently reported
by Ransom \etal (2003).

\Section{Conclusions}

We have used the archival HST/WFPC2 data to identify 15 new variables
located in fields covering central regions of three nearby globular clusters. 
Among others the sample includes seven W~UMa-type compact binaries, 
one detached eclipsing binary and two SX~Phe stars. We do not have any 
definitive information
about membership status of these stars but note that investigated
regions are strongly dominated by stars belonging to the respective 
clusters. Also the location of detected variables on the color-magnitude 
diagrams favoures the hypothesis that most of them are members of 
investigated  clusters.
It is possible to obtain more firm conclusions about membership
status of detected variables by obtaining second epoch HST/WFPC2 observations
and performing suitable proper motion studies (\eg Anderson et al. 2003). 

Short time coverage and small number of available images limited 
sample of detectable variables to objects with period of the order of few
hours. In fact, all but one of identified objects show continuous 
variability which makes their detection far easier than detection of 
variables with a short duty cycle, such as detached
eclipsing binaries. Another factor limiting completness of derived 
sample of variables even for objects with short periods is a presence 
of many bright stars in the investigated fields. Images of these bright stars 
are severely saturated on the WFPC2 frames. Large number of 
dead/bad pixels on WFPC2 CCD's is yet another limiting factor. 

\Acknow{
We would like to thank Grzegorz Pojma\'nski and Alex Schwarzenberg-Czerny 
for providing some useful software which was used in this project. 
We are grateful to Andrew Dolphin for helpful hints on the HSTphot package.

This work is based on observations with the NASA/ESA Hubble Space
Telescope, obtained from the Data Archive at the Space
Telescope Science Institute, which is operated by the Association
of Universities for Research in Astronomy, Inc., under
NASA contract NAS 5-26555. These observations are associated
with programs \#6121, \#7379 and \#8278.

JK was supported by the Polish KBN grant 5P03D00421.
}

\begin{figure}[htb]
\hglue-0.5cm\psfig{figure=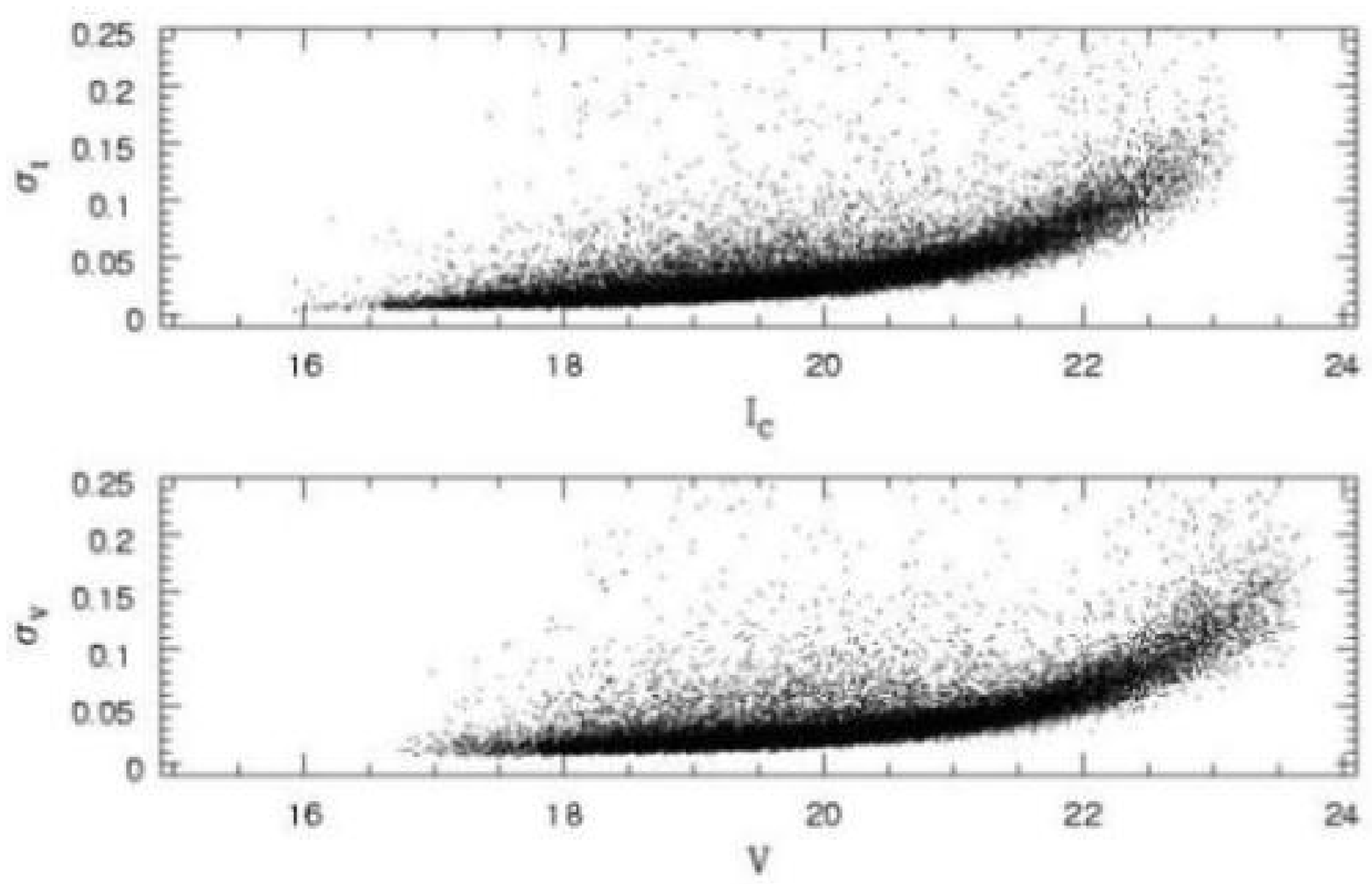,width=13.0cm,angle=0}
\FigCap{
The {\it rms} of derived time-series photometry \vs the
average magnitude for M13 stars.
}
\end{figure}

\begin{figure}[htb]
\hglue-0.5cm\psfig{figure=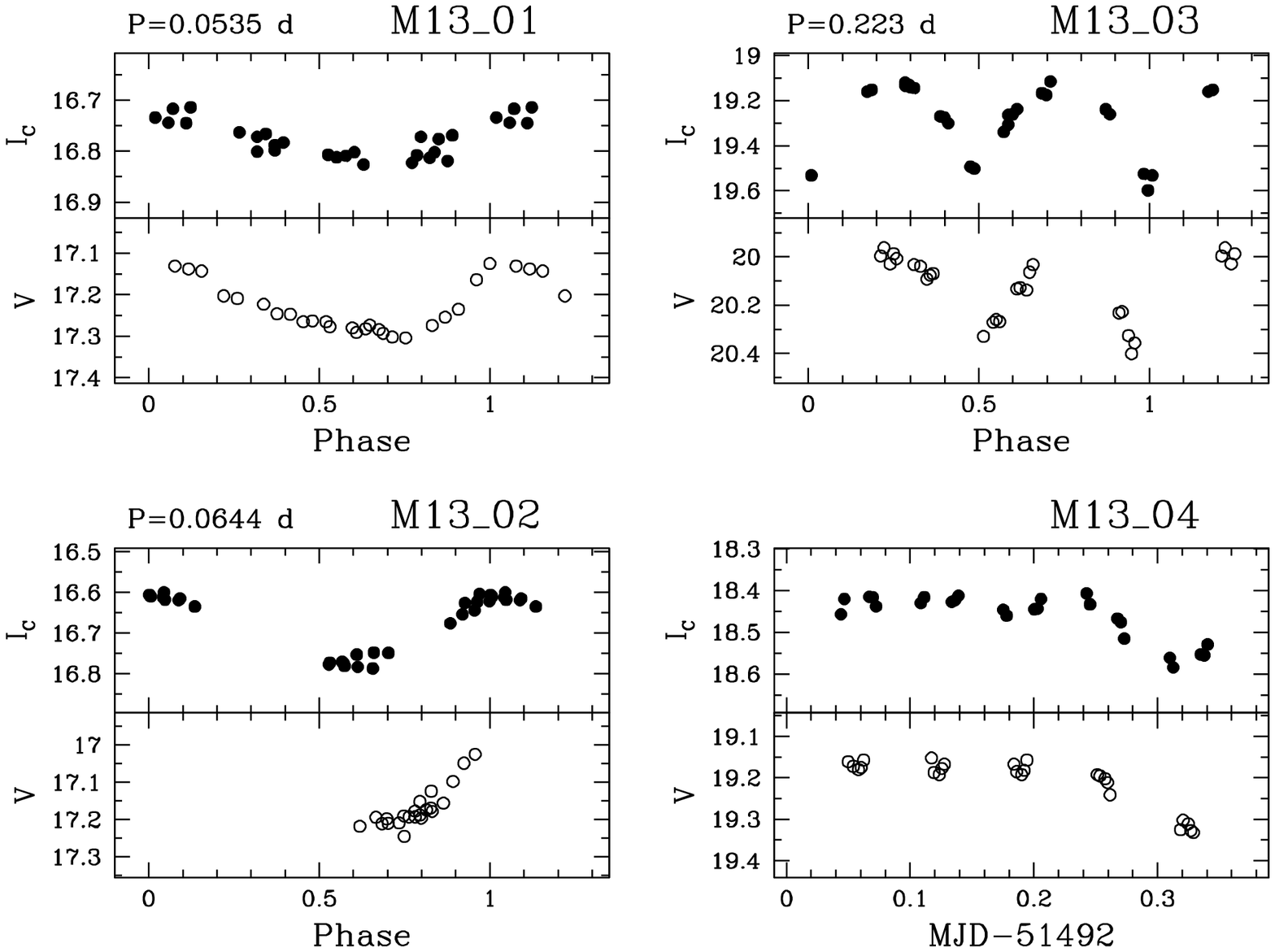,width=13.0cm,angle=0}
\FigCap{
Light curves for newly detected variables from the field of M13.
}
\end{figure}

\begin{figure}[htb]
\hglue-0.5cm\psfig{figure=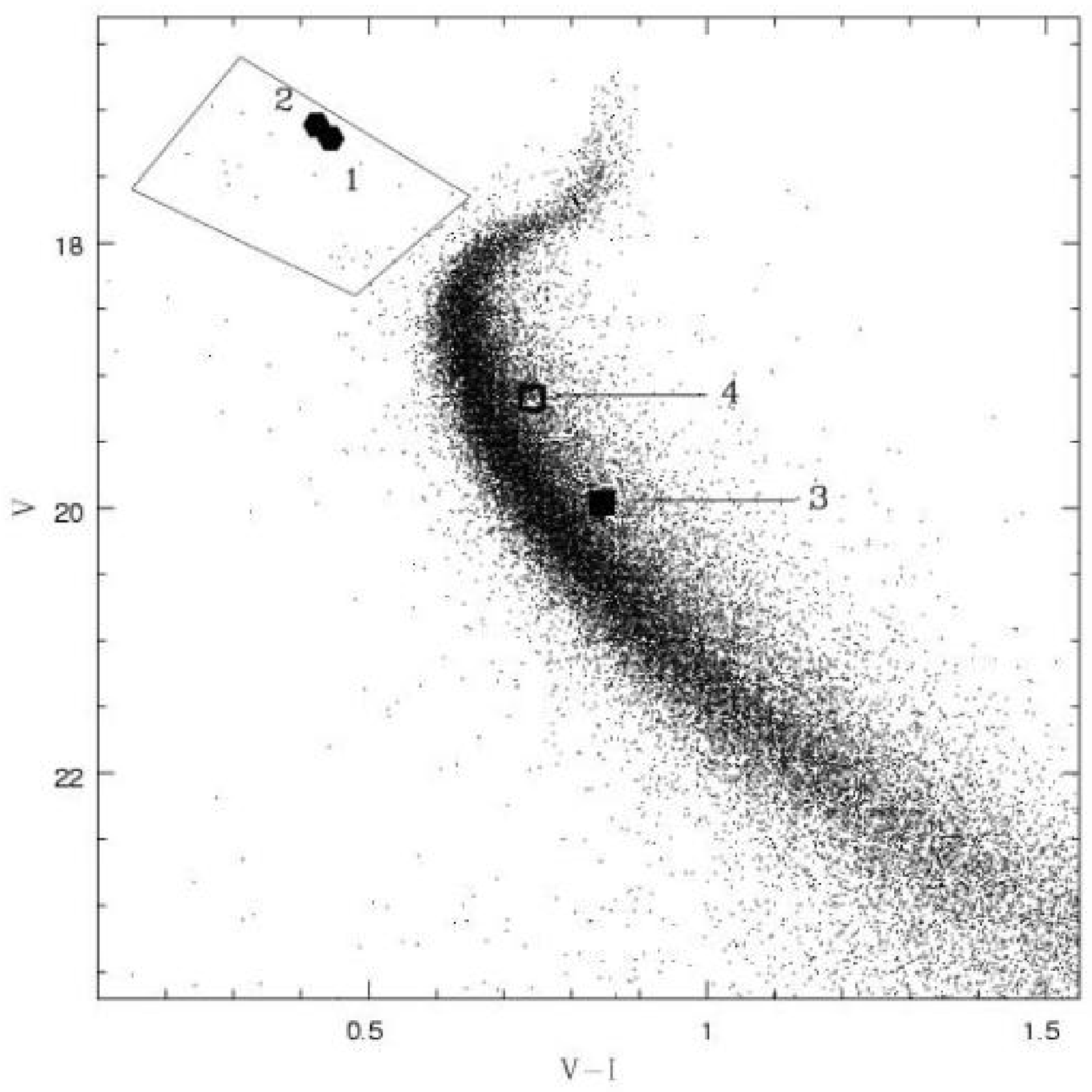,width=13.0cm,angle=0}
\FigCap{
Color-magnitude diagram of M13 with the positions of detected
variables marked.
}
\end{figure}

\begin{figure}[htb]
\hglue-0.5cm\psfig{figure=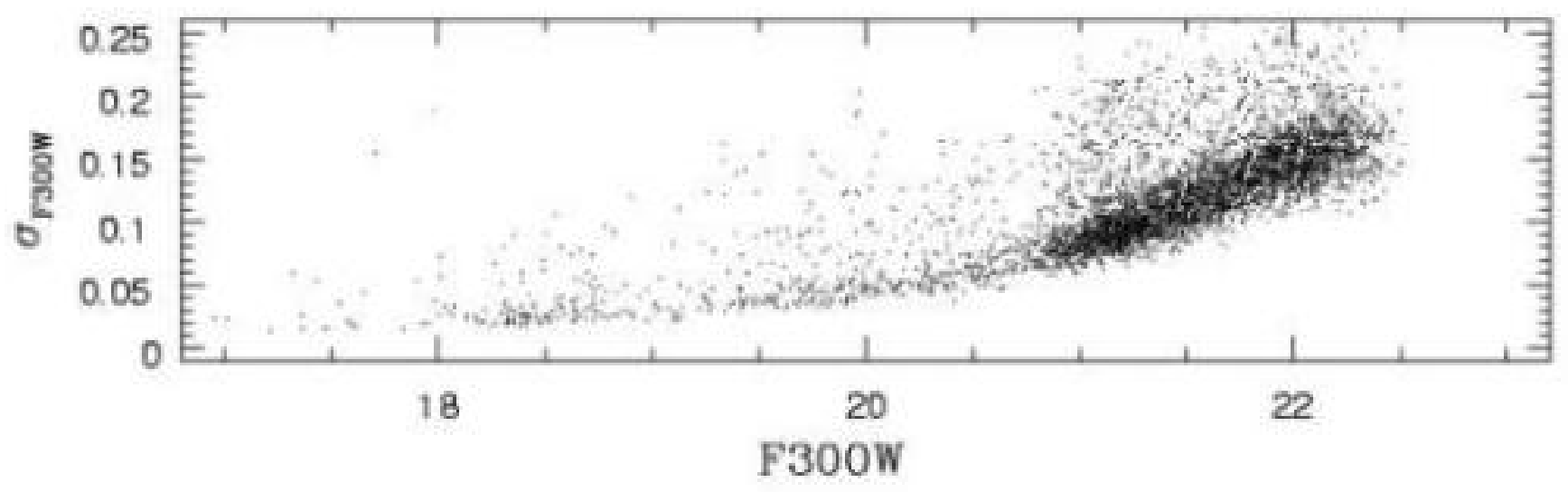,width=13.0cm,angle=0}
\FigCap{
The {\it rms} of derived time-series photometry \vs the average
magnitude for NGC~6712 stars.
}
\end{figure}

\begin{figure}[htb]
\begin{center}
\hglue-0.5cm\psfig{figure=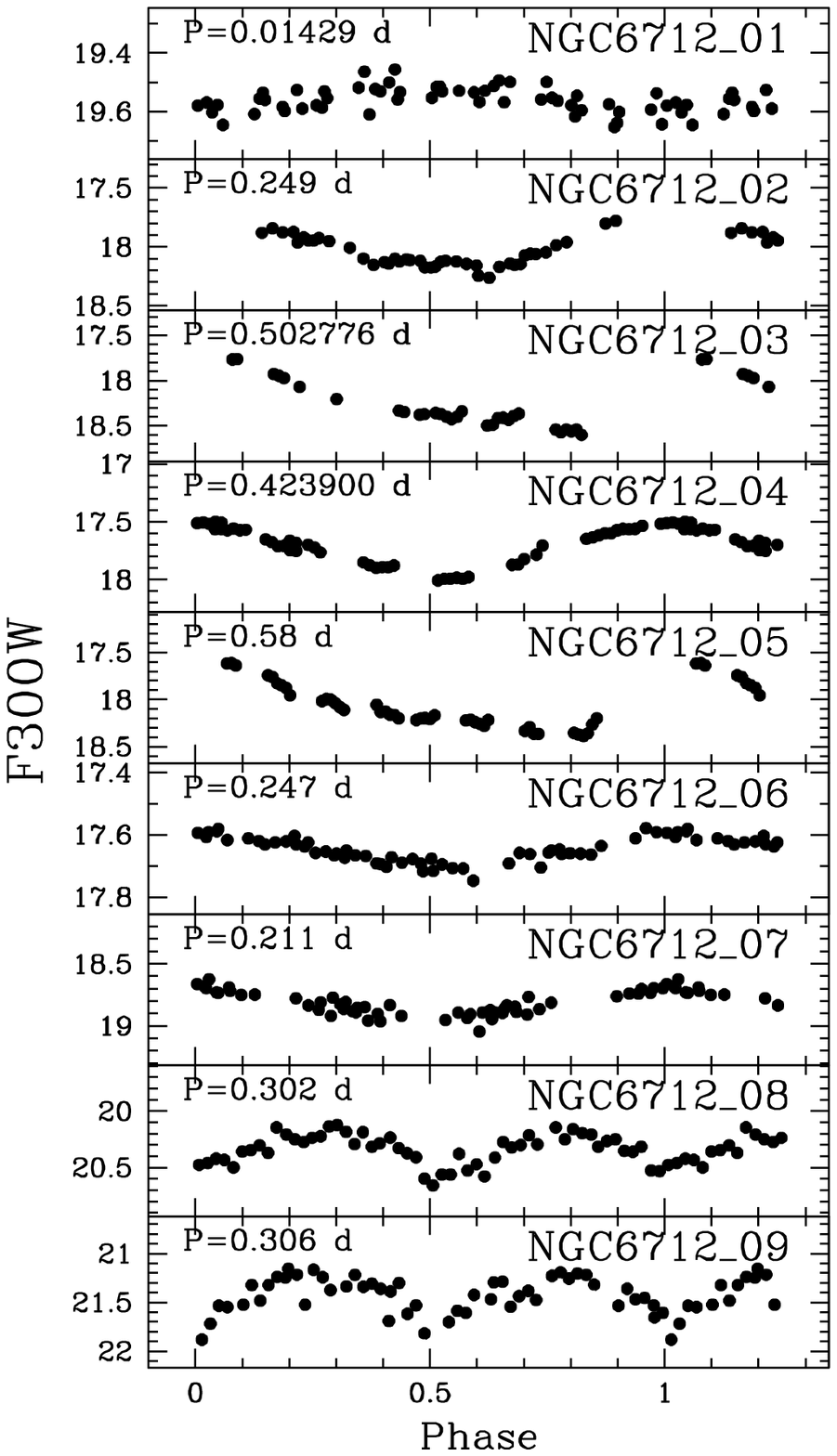,width=7.0cm,angle=0}
\end{center}
\FigCap{
Phased F300W filter light curves for variables detected
in the field of NGC~6712.
}
\end{figure}

\begin{figure}[htb]
\hglue-0.5cm\psfig{figure=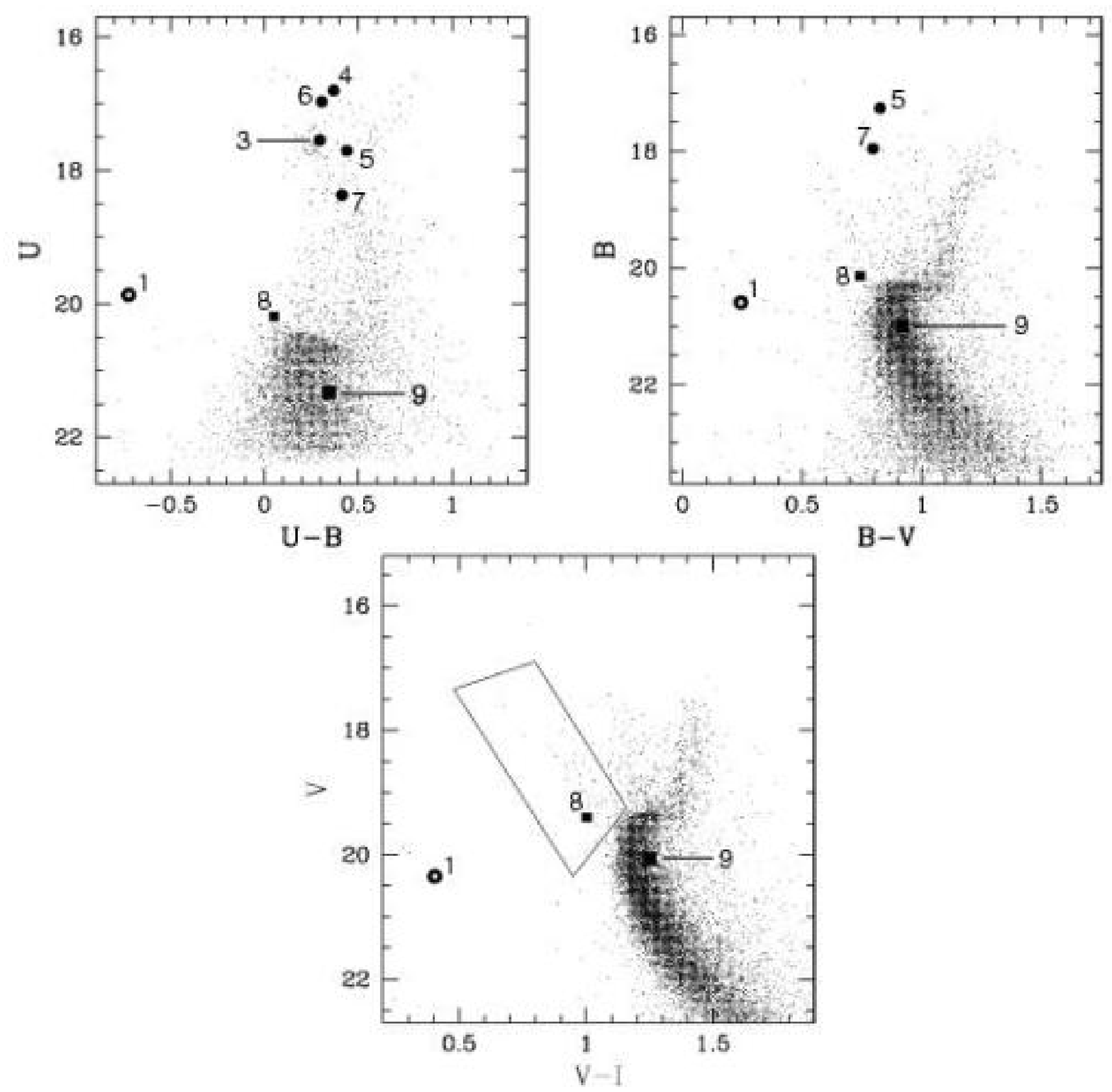,width=13.0cm,angle=0}
\FigCap{
Location of detected variables in the color-magnitude diagrams
of NGC~6712.
}
\end{figure}  

\begin{figure}[htb]
\hglue-0.5cm\psfig{figure=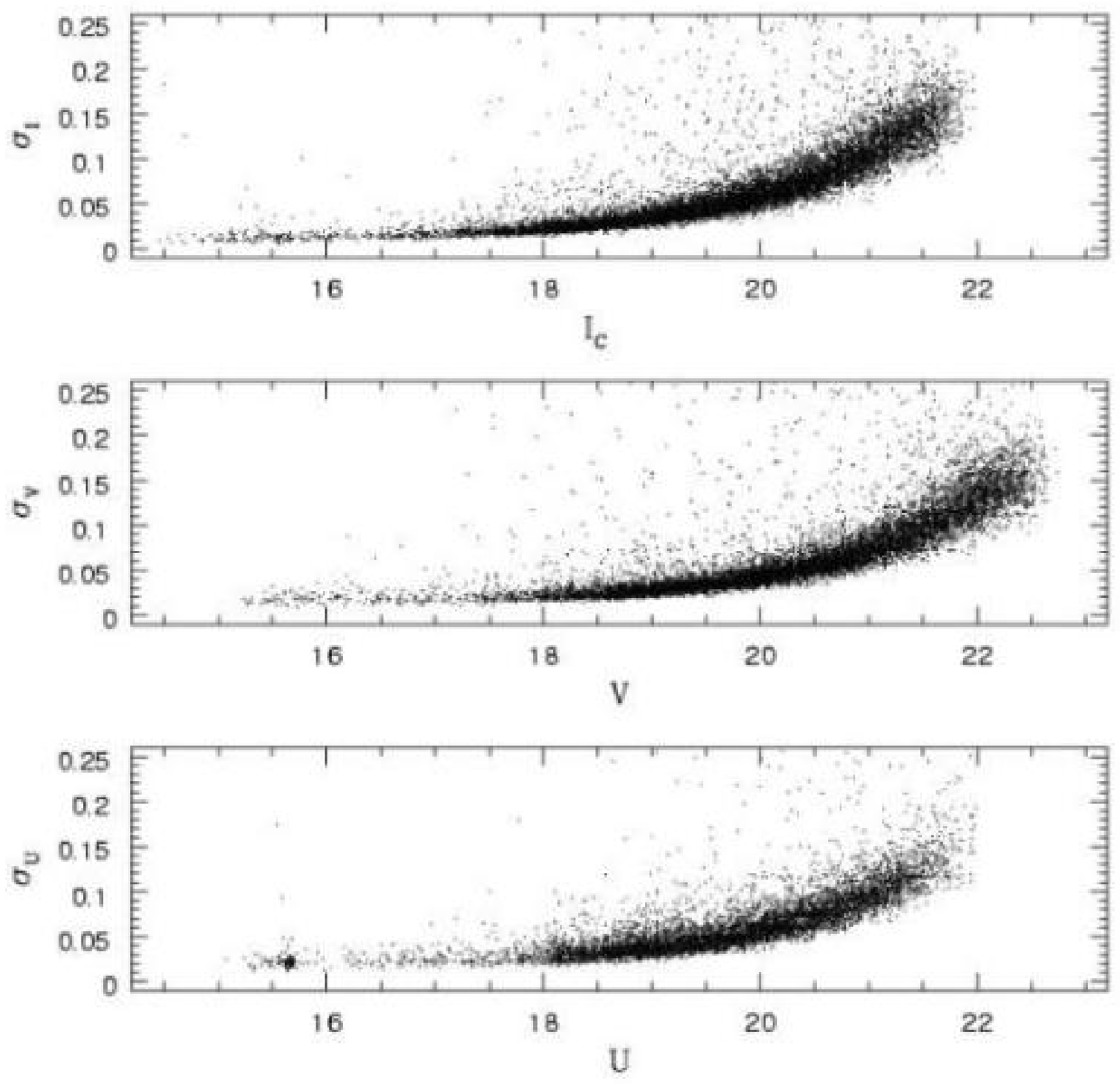,width=13.0cm,angle=0}
\FigCap{
The {\it rms} of derived time-series photometry \vs the
average magnitude for M30 stars.
}
\end{figure}

\begin{figure}[htb]
\hglue-0.5cm\psfig{figure=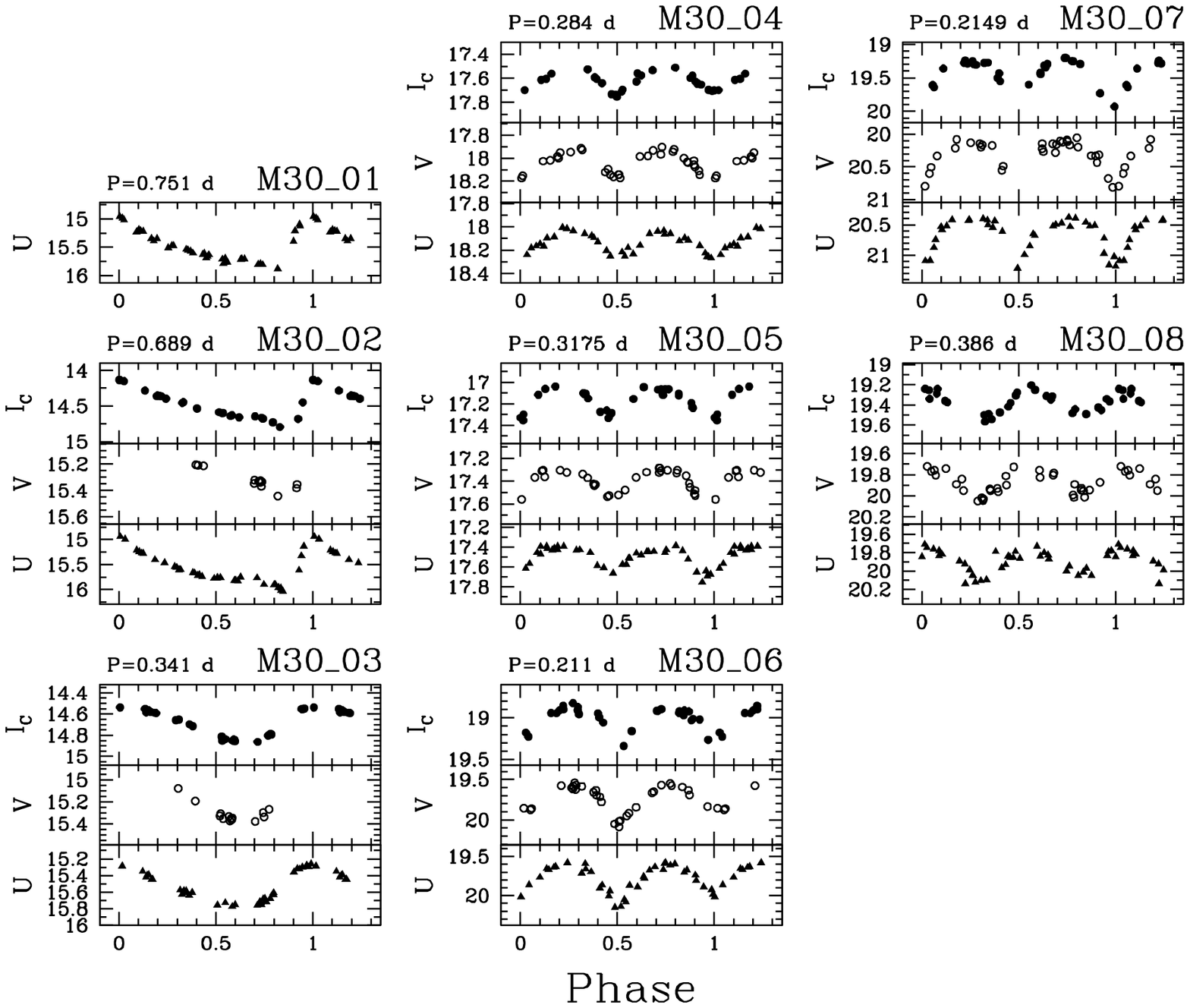,width=13.0cm,angle=0}
\FigCap{
Phased light curves for variables detected in the field of M30.
}
\end{figure}

\begin{figure}[htb]
\hglue-0.5cm\psfig{figure=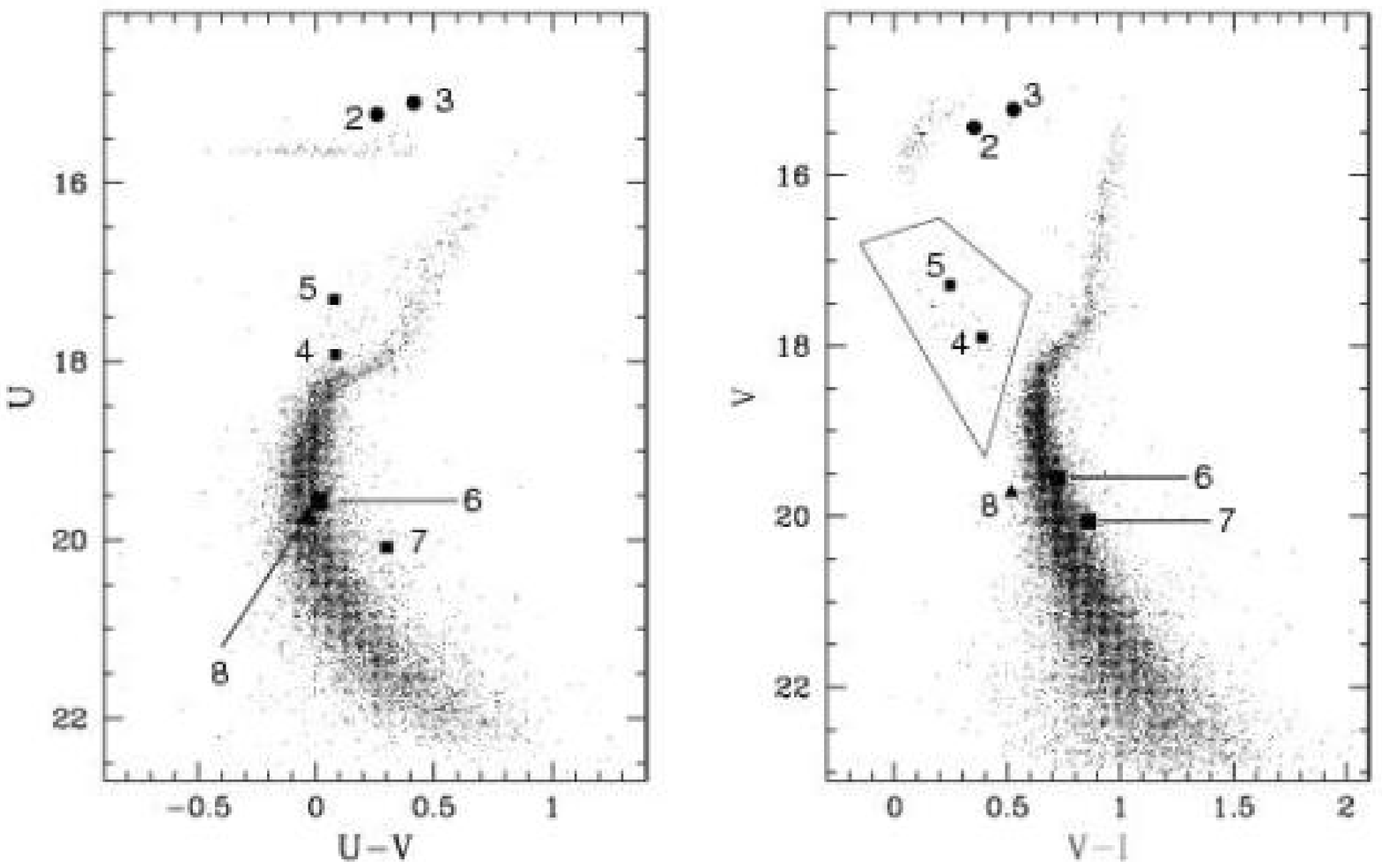,width=13.0cm,angle=0}
\FigCap{
The color-magnitude diagrams of M30 with location of detected 
variables marked.
}
\end{figure}  

\end{document}